\documentclass[fleqn,usenatbib]{mnras}

\usepackage{graphicx}
\usepackage{amssymb}
\usepackage{amsmath}

\usepackage{caption}

\title{A new method of testing the gravitational redshift effect with radio interferometers}

\author[Pilipenko et al.]{\parbox{\textwidth}{
S.~V.~Pilipenko$^{1}$\thanks{E-mail: spilipenko@asc.rssi.ru}, 
D.~A.~Litvinov$^{1,2,3}$,
A.~I.~Filetkin$^2$,
V.~N.~Rudenko$^2$
}
\vspace{0.4cm}\\
\parbox{\textwidth}{
$^1$Astro Space Center of Lebedev Physical Institute, Russian Academy of Sciences, Profsoyuznaja 84/32, 117997 Moscow, Russia\\
$^2$Sternberg Astronomical Institute, Lomonosov Moscow State University,
Universitetsky pr.~13, 119991 Moscow, Russia\\
$^3$Bauman Moscow State Technical University, 2-ya Baumanskaya 5, 105005 Moscow, Russia}
}

\date{Accepted XXX, Received YYY, in original form ZZZ.}

\begin{document}
\maketitle

\begin{abstract}
We propose a new method to measure gravitational redshift effect using simultaneous interferometric observations of a distant radio source to synchronize clocks. The first order by $v/c$ contribution to the signal (the classical Doppler effect) is automatically canceled in our setup. When other contributions from the velocities of the clocks, clock imperfection and atmosphere are properly taken into account, the residual gravitational redshift can be measured with the relative precision of $\sim 10^{-3}$ for RadioAstron space-to-ground interferometer or with precision up to few $10^{-5}$ with the next generation of space radio interferometers.
\end{abstract}


\section{Introduction}

General relativity forms the basis of the modern physical picture of nature.
However, it is not compatible with quantum theory \citep[e.g.][]{carlip-2008-cqg}. Attempts to unify general
relativity and quantum theory lead to violation of the Einstein Equivalence
Principle (EEP). Improving the accuracy of experimental tests of the EEP is thus one of the most important ways of finding hints to new physics \citep{braginsky-rudenko-1970-spu,braginsky-panov-1972-jetp,braginsky-manukin-1977-book,will-2014-lrr}.

One way to experimentally test the EEP is by measuring the gravitational redshift:
\begin{equation}
{\Delta \nu_\mathrm{grav} \over \nu} = {\Delta U \over c^2}
(1 + \epsilon),
\end{equation}
where $\delta U$ is the gravitational potential difference between the measurement
points, $c$ is the speed of light and $\epsilon$ is the EEP violation parameter
to be determined. The most accurate results may be obtained
by maximizing $\Delta U$, i.e. using measurements performed with the help of spacecrafts in the gravitational fields of the Earth, Moon or Sun. In
what follows we assume that one of the measurement points is at the
spacecraft (SC), while the other is at a ground tracking station (TS) located
on the Earth.

There are several options for setting up an experiment to measure the gravitational redshift. In the simplest ``one-way'' experiment, a highly stable and/or accurate  clock on board a spacecraft is used to synchronize its downlink signal frequency, $f$, which is then measured at the receiving station. However, such a scheme is difficult to realize due to the fact that in addition to the gravitational redshift, many other significant factors contribute to the frequency shift: the Doppler effect, frequency shift due to the troposphere and ionosphere, etc.
Instead, a specific configuration of spacecraft's communication links pioneered
by Gravity Probe A \citep{vessot-levine-1979-grg,vessot-levine-1980-prl} is
usually used to cancel the Doppler and propagation effects. In this ``compensation scheme'' the ground TS transmits a signal of frequency $\nu$, which is received by the SC with frequency $\nu'$. The SC then coherently re-transmits this signal back to Earth, and also simultaneously emits a signal of frequency of $\nu$ synchronized to its onboard clock. By comparing the initial frequency and the frequencies of the two signals received at the TS, it is possible to eliminate the tropospheric frequency shift and also that due to the Doppler effect in the first order in $v/c$. This scheme was also successfully used on RadioAstron satellite \citep{biriukov-2014-ar,litvinov-mg14}.

In this paper, we propose a new method to test the gravitational redshift that is based on clock comparison by observing distant quasars with a space-to-ground very-long baseline interferometer (VLBI), such as RadioAstron \citep{kardashev-2013-ar}. The paper
is organized as follows. In Section 2 we give an  outline of the experiment under a simplifying assumption that the TS used to measure the SC's radial velocity
and the ground radio telescope (GRT), where the ground clock
resides, are co-located. In Section 3 we present a detailed account of the experiment taking into
account that the TS and the GRT may be separated by several hundred or even
thousand kilometers. In Section 4 we consider the accuracy of the
proposed approach achievable with the RadioAstron space VLBI mission. We conclude by discussing the results and possible applications
in Section 5.

\section{Simplified model}
\begin{figure}
\centering
\includegraphics[width=8cm]{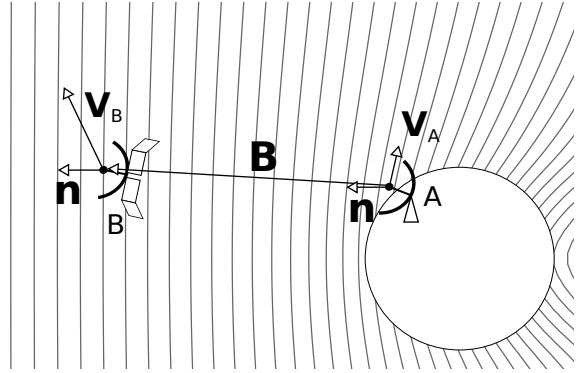}
\caption{The scheme of the experiment to measure the gravitational redshift using a space-to-ground radio interferometer. Vertical lines show wave fronts of the plane wave signal coming from a distant radio source, they are curved near the Earth due to gravitational redshift.}
\label{fig:schema}
\end{figure}

Consider a two-element radio interferometer made up of stations $A$ and $B$, each equipped with a radio telescope, a reference
frequency standard (clock) and a VLBI back-end for signal registration. Suppose
each radio telescope is receiving the same radio signal emitted by a distant point-like radio source, e.g. a quasar. At each station, the signal is heterodyned using a reference signal from its local clock,
usually a hydrogen maser, and then digitized. After the signal has been recorded
at each station, standard VLBI data processing is performed, probably at
a different site, which allows one to determine the delay
 between the stations, $\tau$, by maximizing the so-called visibility function \citep{thompson-moran-swenson-2017-book}. Several factors contribute to this delay: the difference in the optical path lengths  from the source to each stations, delays in the equipment, station clock offsets, gravitational time dilation experienced by the station clocks,
etc. This delay is slowly changing and in the first order may approximated
by:
\begin{equation}
\tau_t = \tau + \dot{\tau}(t - t_0) + ...,
\label{eq:delay_model}
\end{equation}
where $t_0$ is some fixed moment of time. If one of the stations is a spacecraft, the quadratic term is also usually taken into account to compensate for acceleration
of the spacecraft. Let station $A$ be the reference one, then: 
\begin{equation}
\dot{\tau} = {\nu_B - \nu_A \over \nu_A},
\label{eq:freq_shift}
\end{equation}
where $\nu_A$ and $\nu_B$ are the frequencies of the signal received by, respectively, station $A$ and $B$. On the other hand, $\dot{\tau}$ can be
obtained by differentiating the delay:
\begin{equation}
\dot{\tau} = -{\dot{\mathbf{B}} \cdot \mathbf{n} \over c} +
\dot{\tau}_\mathrm{grav} + \dot{\tau}_\mathrm{clock} + ...,
\label{eq:tau_dot}
\end{equation}
where $\dot{\mathbf{B}} = \mathbf{V}_B - \mathbf{V}_A$ is the rate of change
of the baseline vector (i.e. the vector pointing from $A$ to $B$) and $\mathbf{n}$ is a unit vector in the direction of the source (Fig.~\ref{fig:schema}). The first term in this equation corresponds to the Doppler effect, however, for simplicity, it is shown here only in the first order in $v/c$.

Equation \eqref{eq:tau_dot} can be used to measure $\dot{\tau}_\mathrm{grav}$ if the terms on the right-hand side are known or small enough. The ground
station position and velocity are usually known with sufficient accuracy, and the main uncertainty is due to additional phase rotation in the ionosphere and troposphere. Another major difficulty is to determine the velocity of
the spacecraft, let it be station $B$, with sufficient accuracy. Therefore,
apart from the ground station $A$, we must consider another tracking station
(or several such stations), which measures the spacecraft radial velocity by Doppler tracking. We will
assume that this additional tracking station is close enough to station
$A$, so that the projection of the spacecraft velocity onto vector $\mathbf{B}$ can
be measured accurately enough by Doppler tracking at this additional station.
The velocity component in the direction orthogonal to $\mathbf{B}$ can be
be determined with much lower accuracy. This fact restricts
the allowed orientations of the interferometer significantly, making observations possible only in the directions very close to that of vector $\mathrm{B}$.
In this case the transverse velocity  gives only a small contribution
to Eq.~\eqref{eq:tau_dot}.

In this case the distant radio source, the spacecraft and the tracking station are approximately aligned, with the spacecraft being in the middle. Therefore, the Doppler effect is, in the first order, of different sign as measured
by the receiving station compared to Eq.~\eqref{eq:tau_dot_srt1}. In other words, in the disposition of Fig.~\ref{fig:schema} the spacecraft is moving away from the Earth and the tracking station registers a negative Doppler shift,
while the spacecraft is approaching the source and registers a positive Doppler
shift, which will accordingly be reflected in the delay. This means that one can get rid of the Doppler effect in the first order, just like in the compensation scheme.

\section{Accuracy of the experiment}
\subsection{Two stations}
Suppose that during the time of measuring the position of the stations vary slightly compared to the distance between them. Then, based on the equation \eqref{eq:tau_dot}, we obtain the following relation for the measurement error $\dot{\tau}_\mathrm{grav}$:
\begin{multline}
\sigma_\mathrm{grav}^2 = {1 \over c^2} \left( \sigma_{V_A}^2 \cos^2 \alpha + \sigma_{V_B}^2 \cos^2 \alpha + |\mathbf{V}_B - \mathbf{V}_A|^2\sigma_{\cos \alpha}^2 \right)
\\+ \sigma_\mathrm{fit}^2 + \sigma_\mathrm{clock}^2 + \sigma_\mathrm{atm}^2 + ...,
\label{eq:err_grt}
\end{multline}
where $\sigma_{V_A}$, $\sigma_{V_B}$ --- station speed errors, 
$\cos \alpha$ --- cosine of the angle between the base vector and the direction to the source,
$|\mathbf{V}_B - \mathbf{V}_A|$ --- absolute difference of station speeds,
$\sigma_{\cos \alpha} $ --- measurement error of the cosine of the angle between the base vector and the direction to the source,
$\sigma_\mathrm{fit}$ --- error in determining the rate of change of the delay in the process of searching for interference lobes,
$\sigma_\mathrm{clock}$ --- clock error,
$\sigma_\mathrm{atm}$ --- phase change error as it passes through the atmosphere.

The velocity measurement errors for a network of terrestrial radio telescopes participating in geodetic measurements are $\sim$ 1~$\mu$m/s. Direction error to the source $\sigma_{\cos \alpha} = \sigma_\alpha \sin \alpha$, where $\sigma_\alpha\sim 10^{-8}$ for sources used for geodetic tasks \citep{sovers-1998-rmp,thebault-2015-eps,pavlis-2008-jgr}. The accuracy of the clock $\sigma_\mathrm{clock}$ depends on the accumulation time. For estimates, we take the characteristic value of $\sigma_\mathrm{clock} = 3\times 10^{-15}$ for an accumulation time of 1000 seconds. 

The error $\sigma_\mathrm{fit}$ depends on the delay model used when searching for interference lobes, on the algorithm for maximizing the visibility function, and on the signal-to-noise ratio for a particular experiment. We start from the delay model used in the PIMA \cite{Petrov11} program. Namely, the interference lobe search procedure consists in finding the maximum of the function
\begin{multline}
C(\tau_p, \tau_g, \dot{\tau}_p, \dot{\tau}_g) = \sum_k \sum_j c_{kj} w_{kj} \\
\times 
e^{i(
\omega_0 \tau_p + 
\omega_0 \dot{\tau}_p (t_k - t_0) + 
(\omega_j - \omega_0) \tau_g +
(\omega_j - \omega_0) \dot{\tau}_g (t_k - t_0)
 )},
\label{eq:delaymodel}
\end{multline}
where $\tau_p$, $\tau_g$ --- phase and group delay, $\dot{\tau}_p$, $\dot{\tau}_g$ --- phase and group delay variation rates, $c_{kj}$ --- discretized cross-correlation function, index $k$ runs on time, index $j$ --- on frequency, $w_{kj}$ --- weight measurements, $\omega_0$ and $t_0$ --- reference circular frequency and time.

The frequency change due to the gravitational redshift leads to a change in both the phase and group rate of change of the delay in accordance with the equation \eqref{eq:freq_shift}. Transforming equation (14) from \citet{Whitney76}, we obtained the accuracy of determining these velocities:
\begin{equation}
  \sigma_{\dot{\tau}_p} = {6 \over \pi} {\eta \over S \nu_0 \Delta t},
  \label{eq:sigma_rate_p}
\end{equation}

\begin{equation}
  \sigma_{\dot{\tau}_g} = {6 \over \pi} {\eta \over S \Delta \nu \Delta t},
  \label{eq:sigma_rate_g}
\end{equation}
where $\eta$ is the quantization efficiency coefficient, $S$ is the signal-to-noise ratio, defined as the ratio of the antenna temperature to the system noise temperature in the case of identical antennas, $\nu_0 = \omega_0 / 2 \pi $ --- reference frequency, $\Delta \nu $ --- the width of the observation bandwidth, $\Delta t$ --- the accumulation duration. Since radioastronomical observations at frequencies above gigahertz are usually $\nu_0 \gg \Delta \nu$, as can be seen from the equations \eqref{eq:sigma_rate_p} and \eqref {eq:sigma_rate_g}, the accuracy of measuring the phase rate of change of the delay higher than the group, so it is proposed to use $\dot{\tau}_p$  to measure the gravitational redshift.

The estimates of the terms of the equation~\eqref{eq:err_grt} are given in Table~\ref{tab:grt}. The greatest contribution to the total error is given by the station speed errors.
\begin{table*}
\captionsetup{width=.60\textwidth}
\centering
\caption{Estimation of gravitational redshift measurement errors for two radio telescopes in the equation (\ref{eq:err_grt}).}
\begin{tabular}{l|l|l}
\hline
Component & Value & Comment \\
\hline
$\sigma_{V_A}/c$, $\sigma_{V_B}/c$ & $3\times10^{-14}$ & Accuracy 0.01 mm/s \\
$|\mathbf{V}_B - \mathbf{V}_A|\sigma_{\cos \alpha}/c$ & $3\times10^{-15}$ & $\alpha\sim$~$90^\circ$, $\sigma_{\cos \alpha}=10^{-8}$, $|\mathbf{V}_B - \mathbf{V}_A|=100$ ì/ñ \\
$\sigma_\mathrm{fit}$ & $3\times10^{-15}$ & $S=100$, $\nu_0=5$~GHz, $\eta=0.7$, $\Delta t = 1000$~s \\
$\sigma_\mathrm{clock}$ & $3\times10^{-15}$ & $\Delta t = 1000$~s \\
\hline
$\sigma_\mathrm{grav}$ & $5\times10^{-14}$ & \\
\hline
\end{tabular}
\label{tab:grt}
\end{table*}

\subsection{Earth-space interferometer}
If one telescope is in space, direct measurement of the radial velocity component using the Doppler effect between the spacecraft and the ground tracking station is possible. In general, the tracking station does not have to coincide with the ground-based telescope, therefore, we introduce into the scheme shown in Fig.~\ref{fig:schema} another station C, located on the surface of the Earth. The expression for the measurement error of the gravitational redshift is given in Appendix A:
\begin{multline}
\sigma_\mathrm{grav}^2 =
\sigma_{\Delta \nu/\nu}^2 \cos^2 \beta \;+ 
\left( {\nu_B - \nu_C \over \nu_C} \sigma_\beta \sin \beta \right)^2 +
{\sigma_{\Delta v_t}^2 \over c^2 } \sin^2 \beta
\\ +
{\sigma_{V_C}^2 + \sigma_{V_A}^2 \over c^2} \cos^2 \gamma 
+ \left( { |\mathbf{V}_C - \mathbf{V}_A| \over c } \sigma_\gamma \sin \gamma \right) +
\sigma_\mathrm{fit}^2 
\\+ \sigma_\mathrm{clock}^2 + \sigma_\mathrm{atm}^2 + ...,
\label{eq:sigma-3-stations}
\end{multline}
where $\sigma_{\Delta\nu / \nu}$ is the measurement error of the frequency shift of the signal transmitted from the spacecraft to the receiving station C, $\beta$ is the angle between the direction to the source and the CB line connecting the spacecraft and the receiving station, $\gamma$ --- the angle between the direction to the source and the line CA connecting the ground-based telescope and the receiving station, $\sigma_\beta$, $\sigma_\gamma$ are the errors in determining the corresponding angles, $\sigma_{\Delta v_t} $ is the error in measuring the projection of the velocity of spacecraft relative to station C perpendicular to the direction to this station. The remaining quantities were entered in the equation (\ref{eq:err_grt}).

As an example, we consider the space-space interferometer Radioastron. For it, the frequency shift can be measured with an accuracy of about $\sigma _{\Delta\nu / \nu} \sim 3 \times 10^{-13}$, while the lateral velocity is measured by restoring the orbit from many measurements and has an error about 2 mm / s \citep{zaslavsky-2016-la}, which corresponds to the error ${\sigma_{\Delta v_t} \over c} \sim 7 \times 10^{-12} $. Therefore, to increase the accuracy of gravitational redshift measurements, it is necessary to make $\sin\beta $ as small as possible. The table \ref{tab:srt} provides estimates of accuracy for the angle $\beta = 3 ^ \circ$. Even with such a small value of this angle, the measurement error $\dot{\tau}'_\mathrm{grav}$ for the ground-space interferometer is determined mainly by the error in measuring the speed of the spacecraft perpendicular to the direction to the receiving station.

\begin{table*}
\captionsetup{width=.59\textwidth}
\centering
\caption{Estimation of gravitational redshift measurement errors for the ground-space interferometer in the equation (\ref{eq:err-srt}).}
\begin{tabular}{l|l|l}
\hline
Component & Value & Comment \\
\hline
$\sigma_{\Delta \nu/\nu}$ & $3\times10^{-13}$ &  \\
${\nu_B - \nu_C \over \nu_C} \sigma_\beta \sin \beta$ & $2\times10^{-15}$ & ${\nu_B - \nu_C \over \nu_C}=3\times10^{-6}$, $\sigma_\beta = 10^{-8}$, $\beta = 3^\circ$ \\
${\sigma_{\Delta v_t} \over c } \sin \beta $ & $3.5\times10^{-13}$ & $\sigma_{\Delta v_t}=20$~mm/s, $\beta=3^\circ$ \\
${\sigma_{V_{A,C}} \over c} \cos \gamma$ & $3\times10^{-14}$ & \\
${ |\mathbf{V}_C - \mathbf{V}_A| \over c } \sigma_\gamma \sin \gamma$ & $3\times10^{-15}$ \\
$\sigma_\mathrm{fit}$ & $1\times10^{-14}$ & $S=20$, $\nu_0=5$~GHz, $\eta=0.7$, $\Delta t = 1000$~s \\
$\sigma_\mathrm{clock}$ & $3\times10^{-15}$ & $\Delta t = 1000$~s \\
\hline
$\sigma_\mathrm{grav}$ & $5\times10^{-13}$ & \\
\hline
\end{tabular}
\label{tab:srt}
\end{table*}

\section{Measuring the gravitational redshift using RadioAstron}

In this section we evaluate the accuracy of the experiment achievable with
the space radio telescope RadioAstron based on equation
\eqref{eq:tau_dot}. We add several noise terms on its right-hand side. We assume the noise of the on-board clock \citep{vremya-ch-vch-1010} is characterized by the following power spectral density (PSD):
\begin{equation}
F_\mathrm{clock} = 1.5\times10^{-26} + 7.5\times10^{-31} f^{-1}.
\end{equation}
In our simulations we also include the noise of the troposphere, which, following
\citep{keihm-2004,boehm-2006-jgr,hernandez-pajares-2009-jg} we
characterize by the following PSD:
\begin{equation}
F_\mathrm{atm} = 2.8\times 10^{-28} f^{-0.4}.
\end{equation}
Other parameters of the experimental setup are summarized in Table~\ref{tab:ra-experiment}.
To account for orbit determination errors, we also introduce an uncorrelated gaussian process with zero mean and a standard deviation corresponding to
the magnitude of the velocity error. We assume the duration of a single observation is 2400~s, the experiment lasts for 200 days and consists of 400 observations equally spaced over this time period. 

\begin{figure*}
\centering
\includegraphics[width=16.8cm]{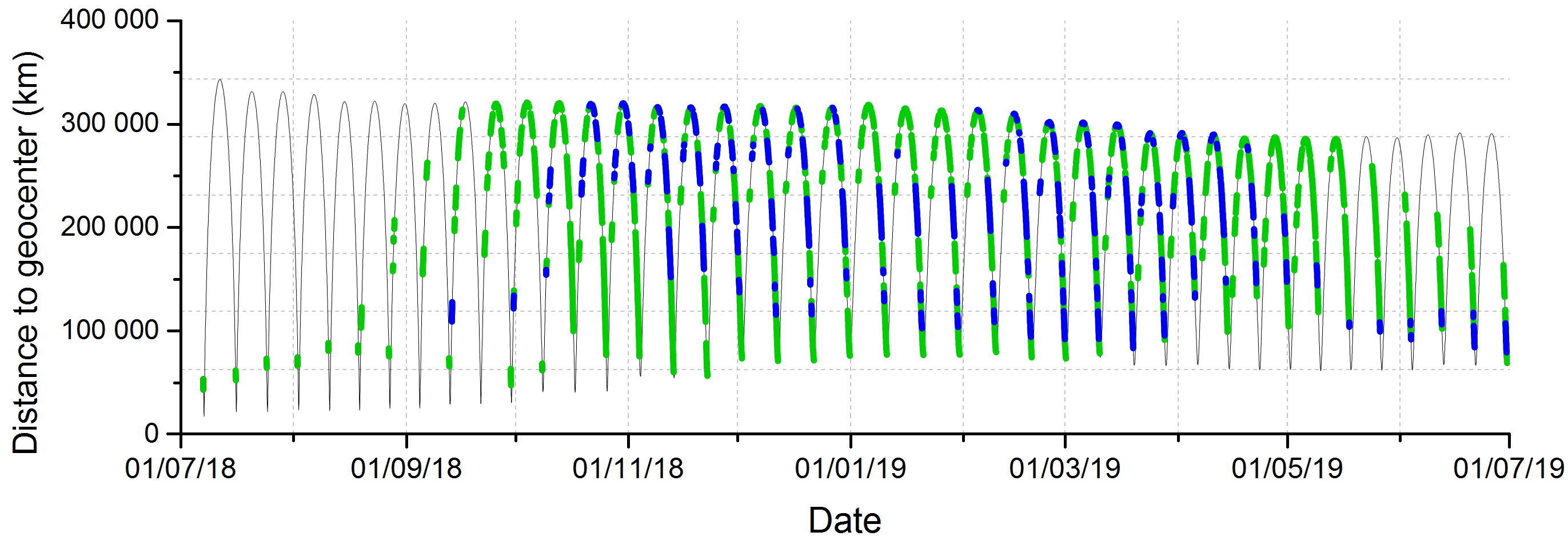}
\caption{The accuracy of the experiment to measure the gravitational redshift using the RadioAstron space-to-ground radio interferometer. Spacecraft velocity determination: 2 mm/s.}
\label{fig:ra-obs-opportunities-ao6}
\end{figure*}

We note that the RadioAstron spacecraft has several observational limitations
(i.e. the spacecraft has attitude limitations with regard to the Earth, the
Sun and the Moon). Therefore it is possible to make observations only during certain
periods of time. We examined the observational capabilities for the experiment
during the period of July 2018 - June 2019 (RadioAstron AO-6 period). The results are presented in Fig.~\ref{fig:ra-obs-opportunities-ao6}  
 
\begin{figure}
\centering
\includegraphics[width=8cm]{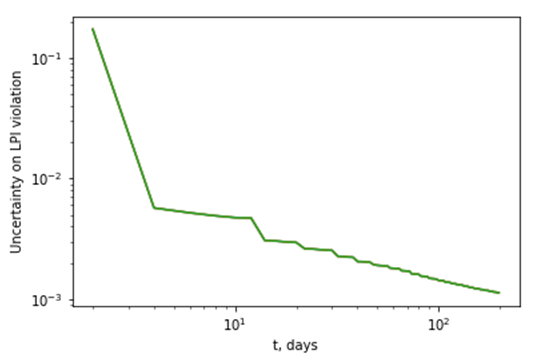}
\caption{The accuracy of the experiment to measure the gravitational redshift using the RadioAstron space-to-ground radio interferometer. Spacecraft velocity determination: 2 mm/s.}
\label{fig:ra-accuracy-2mm-per-s}
\end{figure}

Using the maximum likelyhood estimation we obtain the dependence
of the accuracy of measuring the LPI violation parameter on the duration
of the experiment depicted in Fig.~\ref{fig:ra-accuracy-2mm-per-s}. The accuracy
is mostly limited by the uncertainty of orbit determination, and, specifically,
the radial velocity of the spacecraft relative to the ground tracking station.

\begin{figure}
\centering
\includegraphics[width=8cm]{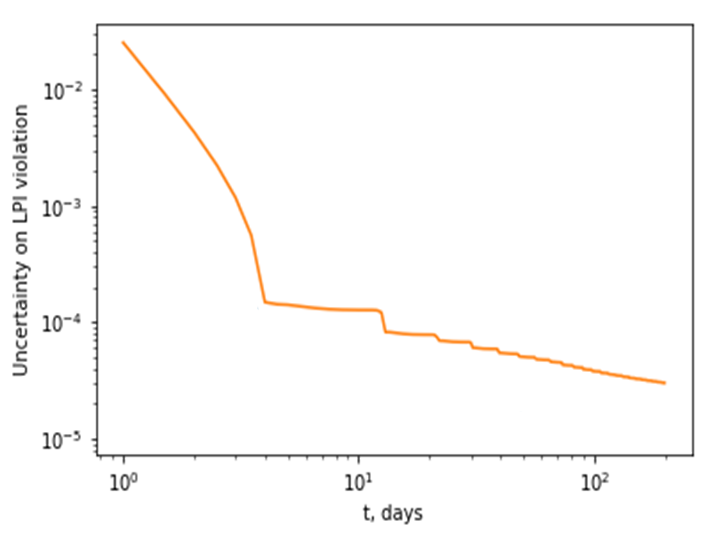}
\caption{The accuracy of the experiment to measure the gravitational redshift using the RadioAstron space-to-ground radio interferometer. Spacecraft velocity determination: 1 $\mu$m/s.}
\label{fig:ra-accuracy-1um-per-s}
\end{figure}

For the purpose of comparison we also estimate the accuracy achievable for
the hypothetical case of the uncertainty of the radial velocity at the level
of 1~$\mu$m/s. We expect this accuracy to be realistic for future space VLBI
missions. In this case the accuracy of the redshift test quickly exceeds that of
of the dedicated Gravity Probe A experiment \cite{vessot-levine-1980-prl} and after
200 days reaches $\sim3\times10^{-5}$ (see Fig.~\ref{fig:ra-accuracy-1um-per-s}).
 
\begin{table}
\captionsetup{width=.3\textwidth}
\centering
\caption{Parameters of the simulation to estimate the accuracy of the experiment
with RadioAstron}
\begin{tabular}{|l|l|l|}
\hline
Parameter & Value \\
\hline
radial velocity accuracy & 2 mm/s \\
\hline
observation frequency & 4.8 GHz\\
\hline
bandwidth & 32 MHz\\
\hline
amplitude quantization & 2 bit\\
\hline
signal-to-noise ratio & 20\\
\hline
\end{tabular}
\label{tab:ra-experiment}
\end{table}

\section{Conclusions}
Using ground-to-space radio interferometers to measure the gravitational redshift effect
appears to be a viable option to testing the Einstein Eqivalence Principle.
We discussed the experiment setup and particularly noted the necessity to
accurately determine the velocity of the spacecraft and discussed a possible
way to decrease the influence of insufficient accuracy of determination of
the transverse component of the velocity by observing sources near the line
of sight of the distant quasar. We showed that the accuracy achievable with the current space VLBI mission RadioAstron is two orders of magnitude below
the current best test of the redshift obtained be Gravity Probe A. If future
space VLBI missions are equipped with better means for orbit determination
(e.g. permanently reachable laser retroreflectors, GNSS receivers and so
on) testing the gravitational redshift may become an important science
case for future space VLBI missions.

\section*{Acknowledgements}

Research for the RadioAstron gravitational redshift experiment is supported by the Russian Science Foundation grant 17-12-01488.
The RadioAstron project is led by the Astro Space Center of the Lebedev Physical Institute of the Russian Academy of Sciences and the Lavochkin Scientific and Production Association under a contract with the Russian Federal Space Agency, in collaboration with partner organizations in Russia and other countries. 

\section*{Appendix A}
The measured values are the rate of change of the delay between ground station A and space B, $\dot{\tau}$, as well as the radial and tangential (with respect to station C) components of the difference in speed of receiving station C and KA. Let us rewrite the equation for rate of change delays (\ref{eq:tau_dot}) so that it includes these measured values:
\begin{equation}
\dot{\tau} = -{ ((\mathbf{v}_B-\mathbf{v}_C)+(\mathbf{v}_C-\mathbf{v}_A)) \mathbf{n} \over c} +
\dot{\tau}_\mathrm{grav} + \dot{\tau}_\mathrm{clock} + ...\,.
\label{eq:tau_dot_srt1}
\end{equation}

The difference $(\mathbf{v}_B-\mathbf{v}_C)$ is represented as the sum of two components of $\Delta v_r$ and $\Delta v_t$, directed along and across the CB line, respectively. Denote the angle between the direction to the source and the line CB as $\beta$, and between the line CA and the direction to the source as $\gamma$. 
Then the equation (\ref{eq:tau_dot_srt1}) is rewritten as:
\begin{equation}
\dot{\tau} = -{ \Delta v_r \cos \beta + \Delta v_t \sin \beta + |\mathbf{v}_C-\mathbf{v}_A| \cos \gamma \over c} +
\dot{\tau}_\mathrm{grav} + \dot{\tau}_\mathrm{clock} + ...\,.
\label{eq:tau_dot_srt2}
\end{equation}

To determine $\Delta v_r$, measurements of the frequency shift of the signal transmitted from the spacecraft to the receiving station C will be used:
\begin{equation}
{\nu_B - \nu_C \over \nu_C} = {\Delta v_r \over c} + \left( {\Delta \nu_{BC} \over \nu} \right)_{c^2}
+ \left( {\Delta \nu_{BC} \over \nu} \right)_\mathrm{grav},
\label{eq:nubc}
\end{equation}
where $\left( {\Delta \nu_{BC} \over \nu} \right)_{c^2}$ is the part of the frequency shift that contains terms of the second and higher orders in $ v/c$ for the effect Doppler, $\left( {\Delta \nu_{BC} \over \nu} \right)_\mathrm{grav}$ is the contribution of the gravitational red shift between the spacecraft and station C to the measured frequency difference. If the reference signal is emitted from the Earth to measure speed, and then the SC is coherently re-emitted to Earth (analogous to the ``Coherent'' mode in Radio Astron), $\left( {\Delta \nu_{BC} \over \nu} \right)_\mathrm{grav}=0$ with accuracy up to the second order in $v/c$. 

Substituting $\Delta v_r$ from (\ref{eq:nubc}) into (\ref{eq:tau_dot_srt2}), we get the following expression:

\begin{equation}
\begin{array}{l}
\dot{\tau} = - {\nu_B - \nu_C \over \nu_C} \cos \beta - {\Delta v_t \over c} \sin \beta - {|\mathbf{v}_C-\mathbf{v}_A| \over c} \cos \gamma + \\
+ \left[\left( {\Delta \nu_{BC} \over \nu} \right)_{c^2} + \left( {\Delta \nu_{BC} \over \nu} \right)_\mathrm{grav}\right] \cos \beta + \dot{\tau}_\mathrm{grav} + \dot{\tau}_\mathrm{clock} + ...\,.
\end{array}
\label{eq:tau_dot_srt3}
\end{equation}

We also introduce the notation
\begin{equation}
\dot{\tau}'_\mathrm{grav} = \dot{\tau}_\mathrm{grav} + \left[\left( {\Delta \nu_{BC} \over \nu} \right)_{c^2} + \left( {\Delta \nu_{BC} \over \nu} \right)_\mathrm{grav}\right] \cos \beta.
\end{equation}
This value depends on the gravitational potentials at points A, B, C, and also contains terms of the second and higher orders in $v/c$ for the Doppler effect. Measuring the value of $\dot{\tau}'_\mathrm{grav}$ allows you to check the Einstein equivalence principle, but the type of its dependence on the potentials of the solar system bodies and speeds of the spacecraft and two ground stations is determined by the model deviation from the Einstein equivalence principle. For example, in the simplest case, the gravitational redshift between two points is described by the equation
\begin{equation}
{\Delta \nu_\mathrm{grav} \over \nu} = {\Delta U \over c^2} (1+\epsilon),
\end{equation}
where $\epsilon$ --- a parameter that characterizes the violation, and it is the same for all navigational bodies. Then
\begin{equation}
\dot{\tau}'_\mathrm{grav} = {U_B(1+\cos \beta) - U_A - U_C \cos \beta \over c^2} (1+\epsilon) + O((v/c)^2),
\end{equation}
where $U_A$, $U_B$, $U_C$ is the sum of the gravitational potentials of all bodies at points A, B and C, respectively, $O((v/c)^2)$ are members of the second and higher orders for the Doppler effect between the spacecraft and the receiving station C.

As a result, by imposing small deviations on all quantities in (\ref{eq:tau_dot_srt3}), we obtain the formula for the measurement error of $\dot {\tau}'_\mathrm{grav}$ for the ground-space interferometer (\ref{eq:sigma-3-stations}).

\bibliographystyle{mnras}

\bibliography{grav}

\begin{thebibliography}{}
\makeatletter
\relax
\def\mn@urlcharsother{\let\do\@makeother \do\$\do\&\do\#\do\^\do\_\do\%\do\~}
\def\mn@doi{\begingroup\mn@urlcharsother \@ifnextchar [ {\mn@doi@}
  {\mn@doi@[]}}
\def\mn@doi@[#1]#2{\def\@tempa{#1}\ifx\@tempa\@empty \href
  {http://dx.doi.org/#2} {doi:#2}\else \href {http://dx.doi.org/#2} {#1}\fi
  \endgroup}
\def\mn@eprint#1#2{\mn@eprint@#1:#2::\@nil}
\def\mn@eprint@arXiv#1{\href {http://arxiv.org/abs/#1} {{\tt arXiv:#1}}}
\def\mn@eprint@dblp#1{\href {http://dblp.uni-trier.de/rec/bibtex/#1.xml}
  {dblp:#1}}
\def\mn@eprint@#1:#2:#3:#4\@nil{\def\@tempa {#1}\def\@tempb {#2}\def\@tempc
  {#3}\ifx \@tempc \@empty \let \@tempc \@tempb \let \@tempb \@tempa \fi \ifx
  \@tempb \@empty \def\@tempb {arXiv}\fi \@ifundefined
  {mn@eprint@\@tempb}{\@tempb:\@tempc}{\expandafter \expandafter \csname
  mn@eprint@\@tempb\endcsname \expandafter{\@tempc}}}

\bibitem[\protect\citeauthoryear{Biriukov, Kauts, Kulagin, Litvinov  \&
  Rudenko}{Biriukov et~al.}{2014}]{biriukov-2014-ar}
Biriukov A.~V.,  Kauts V.~L.,  Kulagin V.~V.,  Litvinov D.~A.,   Rudenko V.~N.,
   2014, \mn@doi [Astronomy Reports] {10.1134/S1063772914110018}, 58, 783

\bibitem[\protect\citeauthoryear{Boehm, Werl  \& Schuh}{Boehm
  et~al.}{2006}]{boehm-2006-jgr}
Boehm J.,  Werl B.,   Schuh H.,  2006, \mn@doi [Journal of Geophysical
  Research: Solid Earth] {10.1029/2005JB003629}, 111, n/a

\bibitem[\protect\citeauthoryear{{Braginskii} \& {Manukin}}{{Braginskii} \&
  {Manukin}}{1977}]{braginsky-manukin-1977-book}
{Braginskii} V.~B.,  {Manukin} A.~B.,  1977, {Measurement of weak forces in
  physics experiments}

\bibitem[\protect\citeauthoryear{{Braginski{\u i}} \& {Rudenko}}{{Braginski{\u
  i}} \& {Rudenko}}{1970}]{braginsky-rudenko-1970-spu}
{Braginski{\u i}} V.~B.,  {Rudenko} V.~N.,  1970, \mn@doi [Soviet Physics
  Uspekhi] {10.1070/PU1970v013n02ABEH004204}, \href
  {http://adsabs.harvard.edu/abs/1970SvPhU..13..165B} {13, 165}

\bibitem[\protect\citeauthoryear{{Braginski{\v i}} \& {Panov}}{{Braginski{\v
  i}} \& {Panov}}{1972}]{braginsky-panov-1972-jetp}
{Braginski{\v i}} V.~B.,  {Panov} V.~I.,  1972, Soviet Journal of Experimental
  and Theoretical Physics, \href
  {http://adsabs.harvard.edu/abs/1972JETP...34..463B} {34, 463}

\bibitem[\protect\citeauthoryear{Carlip}{Carlip}{2008}]{carlip-2008-cqg}
Carlip S.,  2008, Classical and Quantum Gravity, 25, 154010

\bibitem[\protect\citeauthoryear{Hern{\'a}ndez-Pajares
  et~al.,}{Hern{\'a}ndez-Pajares et~al.}{2009}]{hernandez-pajares-2009-jg}
Hern{\'a}ndez-Pajares M.,  et~al., 2009, \mn@doi [Journal of Geodesy]
  {10.1007/s00190-008-0266-1}, 83, 263

\bibitem[\protect\citeauthoryear{Kardashev et~al.,}{Kardashev
  et~al.}{2013}]{kardashev-2013-ar}
Kardashev N.~S.,  et~al., 2013, \mn@doi [Astronomy Reports]
  {10.1134/S1063772913030025}, 57, 153

\bibitem[\protect\citeauthoryear{{Keihm}, {Tanner}  \& {Rosenberger}}{{Keihm}
  et~al.}{2004}]{keihm-2004}
{Keihm} S.~J.,  {Tanner} A.,   {Rosenberger} H.,  2004, Interplanetary Network
  Progress Report, \href {http://adsabs.harvard.edu/abs/2004IPNPR.158A...1K}
  {158, 1}

\bibitem[\protect\citeauthoryear{{Litvinov} et~al.,}{{Litvinov}
  et~al.}{2016}]{litvinov-mg14}
{Litvinov} D.~A.,  et~al., 2016, preprint, \href
  {http://adsabs.harvard.edu/abs/2016arXiv160505832L} {} (\mn@eprint {arXiv}
  {1605.05832})

\bibitem[\protect\citeauthoryear{Pavlis, Holmes, Kenyon  \& Factor}{Pavlis
  et~al.}{2012}]{pavlis-2008-jgr}
Pavlis N.~K.,  Holmes S.~A.,  Kenyon S.~C.,   Factor J.~K.,  2012, \mn@doi
  [Journal of Geophysical Research: Solid Earth] {10.1029/2011JB008916}, 117,
  n/a

\bibitem[\protect\citeauthoryear{{Petrov}, {Kovalev}, {Fomalont}  \&
  {Gordon}}{{Petrov} et~al.}{2011}]{Petrov11}
{Petrov} L.,  {Kovalev} Y.~Y.,  {Fomalont} E.~B.,   {Gordon} D.,  2011, \mn@doi
  [\aj] {10.1088/0004-6256/142/2/35}, \href
  {http://adsabs.harvard.edu/abs/2011AJ....142...35P} {142, 35}

\bibitem[\protect\citeauthoryear{Sovers, Fanselow  \& Jacobs}{Sovers
  et~al.}{1998}]{sovers-1998-rmp}
Sovers O.~J.,  Fanselow J.~L.,   Jacobs C.~S.,  1998, \mn@doi [Rev. Mod. Phys.]
  {10.1103/RevModPhys.70.1393}, 70, 1393

\bibitem[\protect\citeauthoryear{Th{\'e}bault et~al.,}{Th{\'e}bault
  et~al.}{2015}]{thebault-2015-eps}
Th{\'e}bault E.,  et~al., 2015, \mn@doi [Earth, Planets and Space]
  {10.1186/s40623-015-0228-9}, 67, 79

\bibitem[\protect\citeauthoryear{{Thompson}, {Moran}  \& {Swenson}}{{Thompson}
  et~al.}{2017}]{thompson-moran-swenson-2017-book}
{Thompson} A.~R.,  {Moran} J.~M.,   {Swenson} Jr. G.~W.,  2017, {Interferometry
  and Synthesis in Radio Astronomy, 3rd Edition},
  \mn@doi{10.1007/978-3-319-44431-4.
}

\bibitem[\protect\citeauthoryear{{Vessot} \& {Levine}}{{Vessot} \&
  {Levine}}{1979}]{vessot-levine-1979-grg}
{Vessot} R.~F.~C.,  {Levine} M.~W.,  1979, \mn@doi [General Relativity and
  Gravitation] {10.1007/BF00759854}, \href
  {http://adsabs.harvard.edu/abs/1979GReGr..10..181V} {10, 181}

\bibitem[\protect\citeauthoryear{{Vessot} et~al.,}{{Vessot}
  et~al.}{1980}]{vessot-levine-1980-prl}
{Vessot} R.~F.~C.,  et~al., 1980, \mn@doi [Physical Review Letters]
  {10.1103/PhysRevLett.45.2081}, \href
  {http://adsabs.harvard.edu/abs/1980PhRvL..45.2081V} {45, 2081}

\bibitem[\protect\citeauthoryear{{Whitney} et~al.,}{{Whitney}
  et~al.}{1976}]{Whitney76}
{Whitney} A.~R.,  et~al., 1976, \mn@doi [Radio Science]
  {10.1029/RS011i005p00421}, \href
  {http://adsabs.harvard.edu/abs/1976RaSc...11..421W} {11, 421}

\bibitem[\protect\citeauthoryear{Will}{Will}{2014}]{will-2014-lrr}
Will C.~M.,  2014, \mn@doi [Living Reviews in Relativity]
  {10.12942/lrr-2014-4}, 17, 4

\bibitem[\protect\citeauthoryear{{Wolf} \& {Blanchet}}{{Wolf} \&
  {Blanchet}}{2016}]{Wolf16}
{Wolf} P.,  {Blanchet} L.,  2016, \mn@doi [Classical and Quantum Gravity]
  {10.1088/0264-9381/33/3/035012}, \href
  {http://adsabs.harvard.edu/abs/2016CQGra..33c5012W} {33, 035012}

\bibitem[\protect\citeauthoryear{Zaslavskiy, Zakhvatkin, Stepanyants, Tuchin
  \& Shishov}{Zaslavskiy et~al.}{2016}]{zaslavsky-2016-la}
Zaslavskiy G.~S.,  Zakhvatkin M.~V.,  Stepanyants V.~A.,  Tuchin A.~G.,
  Shishov V.~A.,  2016, Vestnik NPO im. S.A.~Lavochkina, 3, 25

\bibitem[\protect\citeauthoryear{Zaslavskiy, Zakhvatkin, Stepanyants, Tuchin
  \& Shishov}{vre}{}]{vremya-ch-vch-1010}
{Active on-board hydrogen maser for Radioastron space mission VCH-1010},
  \url{https://www.vremya-ch.com/english/product/index6e49.html?Razdel=8&Id=39}

\makeatother
\end{thebibliography}








\end{document}